\documentclass[aps,prd,floats,nofootinbib,preprintnumbers]{revtex4}

\usepackage[utf8]{inputenc}
\usepackage[dvips]{color,graphicx}
\usepackage{mathrsfs}
\usepackage{amssymb}
\usepackage{amsmath}
\usepackage{feynmp}
\usepackage{slashed}
\usepackage{latexsym}
\usepackage{graphicx}
\usepackage{verbatim}
\usepackage[normalem]{ulem}

\newcommand{\be}{\begin{equation}}
\newcommand{\ee}{\end{equation}}
\newcommand{\ba}{\begin{array}}
\newcommand{\ea}{\end{array}}
\newcommand{\bea}{\begin{eqnarray}}
\newcommand{\eea}{\end{eqnarray}}
\newcommand{\balg}{\begin{align}}
\newcommand{\ealg}{\end{align}}
\newcommand{\bit}{\begin{itemize}}
\newcommand{\eit}{\end{itemize}}
\newcommand{\trm}[1]{\textrm{#1}}
\newcommand{\mbf}[1]{\mathbf{#1}}

\newcommand{\mcl}[1]{\mathcal{#1}}
\newcommand{\mbb}[1]{\mathbb{#1}}
\newcommand{\msc}[1]{\mathscr{#1}}

\newcommand{\Mpc}{\trm{\Mpc}}
\newcommand{\yr}{\trm{\yr}}
\newcommand{\eV}{\trm{\eV}}

\newcommand{\nn}{\nonumber}

\begin{document}

\preprint{NUHEP-TH/14-05}

\title{Non-Unitary Neutrino Propagation from Neutrino Decay}

\author{Jeffrey M. Berryman}
\affiliation{Northwestern University, Department of Physics \& Astronomy, 2145 Sheridan Road, Evanston, IL~60208, USA}

\author{Andr\'e de Gouv\^ea}
\affiliation{Northwestern University, Department of Physics \& Astronomy, 2145 Sheridan Road, Evanston, IL~60208, USA}

\author{Daniel Hern\'{a}ndez}
\affiliation{Northwestern University, Department of Physics \& Astronomy, 2145 Sheridan Road, Evanston, IL~60208, USA}

\author{Roberto L. N. Oliveira}
\affiliation{Northwestern University, Department of Physics \& Astronomy, 2145 Sheridan Road, Evanston, IL~60208, USA}
\affiliation{Instituto de F\'\i sica Gleb Wataghin Universidade Estadual de Campinas, UNICAMP 13083-970, Campinas, S\~ao Paulo, Brasil}

\begin{abstract}

Neutrino propagation in space-time is not constrained to be unitary if very light states -- lighter than the active neutrinos -- exist into which neutrinos may decay. If this is the case, neutrino flavor-change is governed by a handful of extra mixing and ``oscillation'' parameters, including new sources of CP-invariance violation. We compute the  transition probabilities in the two- and three-flavor scenarios and discuss the different phenomenological consequences of the new physics. These are qualitatively different from other sources of unitarity violation discussed in the literature.

\end{abstract}

\maketitle

\setcounter{equation}{0}
\setcounter{footnote}{0}

Neutrino oscillations, first unambiguously observed towards the end of the twentieth century, have proven to be a powerful tool for fundamental physics research. Their observations revolutionized our understanding of neutrinos, revealing that these have tiny but nonzero masses. Moreover, they can be used to reveal new phenomena including the existence of new, weaker-than-weak interactions involving neutrinos and ordinary matter -- which lead to non-standard matter effects \cite{Wolfenstein:1977ue} -- or the existence of light sterile neutrinos or new contributions to the charged-current weak interactions -- which lead to different non-unitary $3\times 3$ leptonic mixing matrices \cite{Lee:1977qz,Lee:1977tib,Appelquist:2003hn,Antusch:2006vwa}.
Neutrino oscillations also provide powerful testbeds of some of the most basic assumptions of fundamental physics \cite{deGouvea:2013onf}, including tests of whether neutrino propagation is Lorentz invariant \cite{Kostelecky:2003cr}, whether neutrinos obey the CPT-theorem \cite{Kostelecky:2003cr,Murayama:2000hm}, whether there are exotic sources of decoherence in the time-evolution of the neutrino states \cite{Lisi:2000prl,Benatti:2000jhep}, etc. 

In this letter, we explore the consequences of the hypothesis that the neutrino propagation Hamiltonian is not Hermitian. When applied to the active neutrinos, these include new parameters for neutrino oscillation observables that are not captured by the different scenarios considered in the literature to date. We compute  the transition probabilities, concentrating on the case where the non-unitary effects are small, and discuss the different qualitative aspects of the associated phenomenology.

It is important to stress that non-unitary time evolution need not be an extravagant hypothesis. Neutrino propagation is non-unitary if one takes into account the possibility that neutrinos interact with and decay into other generic light states; a generalization of what is known to occur in the neutral kaon system \cite{books}. A similar version of this phenomenon is also realized in resonant leptogenesis \cite{Pilaftsis}.

We start by postulating that the neutrino states involved in the production and detection processes are orthonormal. Generically, we refer to these as \emph{flavor eigenstates} $|\nu_{\alpha}\rangle$ and our assumption amounts to imposing 
\be
\langle \nu_\alpha | \nu_\beta \rangle  = \delta_{\alpha\beta} \label{flavor-ort} \,,
\ee
where $\alpha,\beta$ are flavor indices. For the case of the light, active neutrinos of the Standard Model (SM), this assumption certainly holds true if both production and detection occur via the standard charged-current weak interactions \emph{and there are no additional neutrino states}. In that case, $\alpha,\beta=e,\mu,\tau$. In practice, this implies that as long as the production and detection processes occur through the weak interactions, there are no flavor-changing phenomena in the limit that the baseline is much shorter than the dimensionful parameters that govern propagation, as will become clear momentarily. 

It is important to stress the importance of the assumption that there are no additional neutral fermions that can mix with the three SM flavor states. This assumption sets our framework apart from the nonunitarity scenario analyzed in \cite{Antusch:2006vwa} where such an assumption is implicit. In \cite{Antusch:2006vwa}, the neutrino state appearing in the left-handed charged current along with the lepton $\alpha$ is written as the linear combination $|\nu_\alpha\rangle = \sum_i U_{\alpha i}|\nu_i\rangle$ where $i$ runs over all mass eigenstates, including the heavy ones. Since the production of physical  $|\nu_i\rangle$ states is kinematically forbidden if they are sufficiently heavy, the linear transformation that carries from the produced flavor state to the orthogonal mass eigenstates is nonunitary.

On the other hand, we shall see below that in our framework the implicit assumption is that there is new physics which can mediate neutrino decay into as-of-yet-unknown states. The neutrino states involved in weak interactions are those of the SM and they are orthogonal much in the same way as kaon states produced by strong interactions are orthogonal. In flight, however, new physics can produce an effectively \emph{nonunitary propagation}, playing a role analogous to that played by weak interactions in the kaon system.

Neutrino flavor-evolution in space is governed by the usual Schr\"odinger-like equation, valid in the limit of ultra-relativistic neutrinos assuming that the initial states are perfectly coherent,
\be
i\frac{d}{dL}|\nu_\alpha(L) \rangle = [\msc{H}_{\trm{eff}}]_{\alpha\beta}|\nu_\beta(L) \rangle \,, \label{Scheq}
\ee
where $L$ is the distance traversed by the neutrino. The effective Hamiltonian $\msc{H}_{\trm{eff}}$, which we assume is a generic matrix, can be parameterized as 
\be
\msc{H}_{\trm{eff}} = M - i \Gamma,
\label{Heff}
\ee
where $M$ and $\Gamma$ are Hermitian matrices. For $\Gamma = 0$, we have standard oscillations. The eigenvalues of $M$ are, as usual, $m_i^2/2E$, where $E$ is the neutrino energy and $m_i$, $i=1,2,\ldots$, are real. $\Gamma$ also has real eigenvalues and can be diagonalized by a unitary matrix. We explore the most general case where $M$ and $\Gamma$ cannot be simultaneously diagonalized, i.e., the ``mass'' eigenstates need not coincide with the ``decay'' eigenstates.

Time evolution governed by a non-Hermitian Hamiltonian is generically expected if there are new interactions that couple the light neutrinos to new, very light many-particle states. We provide a quick description of the formalism, which has been developed for the kaon system \cite{books,kabir} and can be readily adapted to neutrino propagation. The main difference between the two is that in the neutrino case there are no constraints from CPT invariance which, for the kaons, lead to some simplifications.

Consider a system consisting of light neutrino states $|\nu_{0i}\rangle$ along with new many-particle states $|\phi_{0k}\rangle$, with the index $k$ understood to run over both discrete and continuous labels required to identify such states. It is convenient for this analysis to work in the mass basis so that $|\nu_{0i}\rangle$ and $|\phi_{0k}\rangle$ are eigenstates of a ``free particle'' Hamiltonian $\msc{H}_0$:
\be
\msc{H}_0 |\nu_{0i}\rangle = E_i |\nu_{0i}\rangle \,,\quad \msc{H}_0 |\phi_{0k}\rangle = E(k) |\phi_{0k}\rangle \,.
\ee
The complete propagation Hamiltonian $\msc{H}$ of the system is assumed to involve new interactions and can be split into
\be
\msc{H} = \msc{H}_0 + \msc{H}' .
\ee
In the context of neutrino oscillations, $\msc{H}_0$ describes the standard propagation Hamiltonian for the neutrino mass eigenstates in the absence of new interactions. On the other hand, the new physics piece $\msc{H}'$ is completely general. In particular, it can induce transitions between the $|\nu_{0i}\rangle$ and the $|\phi_{0k}\rangle$.

At any time $t$, the state of the system $|\psi(t)\rangle$ can be written as a linear combination of the light neutrino eigenstates $|\nu_{0i}\rangle$ and the $|\phi_{0k}\rangle$ as
\be
|\psi(t)\rangle = \sum_i c_i(t)|\nu_{0i} \rangle + \sum_k C_k(t) |\phi_{0k}\rangle ,
\ee 
The time evolution of $|\psi(t)\rangle$ is governed by the Schr\"odinger equation
\be
i\frac{d}{dt}\left( \begin{array}{c} 
\mbf{c}(t) \\ \mbf{C}(t) \end{array} \right) = \msc{H} \left( \begin{array}{c} 
\mbf{c}(t) \\ \mbf{C}(t) \end{array} \right) ,\label{exact-eq}
\ee
where $\mbf{c}(t)$ and $\mbf{C}(t)$ are column vectors formed by the coefficients $c_i(t)$ and $C_k(t)$ respectively. Eq.~\eqref{exact-eq} is exact.

Because $\msc{H}$ is Hermitian, the evolution of the complete system is unitary. Any neutrino produced at time zero satisfies $\sum_i |c_i(0)|^2 = 1$,
while the probability that it remains a neutrino at some time $t$ is $P_{\nu\to\nu}=\sum_i |c_i(t)|^2$. It is clear that 
\begin{equation}
P_{\nu\to\nu}=\sum_i |c_i(t)|^2 = 1 - \sum_k |C_k(t)|^2\le 1,
\label{eq:Ple1}
\end{equation}
for all $t$.\footnote{Note that we make no assumptions about the number of $|\nu_{0i}\rangle$ states. New single-particle states -- e.g., sterile neutrinos -- would simply imply that there are more $|\nu_{0i}\rangle$ states than active neutrinos.}

For the case in which the processes of production and detection involve only linear combinations of neutrino states $|\nu_{0i}\rangle$, it has proved useful to devise a way to reduce Eq.~\eqref{exact-eq} to a differential equation only for the the vector $\mbf{c}(t)$. This can be accomplished under the \emph{Weisskopf-Wigner approximation} (WW). WW assumes that the spectrum of accessible $|\phi_{0k}\rangle$ modes is very broad and that the matrix elements of the Hamiltonian with respect to the new states $ \langle\phi_{0j}|\msc{H}'|\phi_{0k}\rangle$ can be neglected. Under these conditions, \cite{kabir, Stodolsky}
\be
i\frac{d}{dt}\,\mbf{c}(t) = \msc{H}_{\trm{eff}}\,\mbf{c}(t) = (M - i\Gamma)\,\mbf{c}(t) , \label{eff-eq}
\ee
where $M$ and $\Gamma$ are Hermitian matrices with matrix elements given by \cite{kabir}
\begin{align}
M_{ij} & = (E_i-\bar{E})\delta_{ij} + \langle\nu_{0i}|\msc{H}'|\nu_{0j}\rangle \nn \\
&\quad - \sum_k \frac{\langle \nu_{0i}|\msc{H}'|\phi_{0k}\rangle\langle \phi_{0k} |\msc{H}'|\nu_{0j}\rangle}{E(k)-\bar{E}} \,,\\
\Gamma_{ij} & = \pi \sum_k \langle \nu_{0i}|\msc{H}'|\phi_{0k}\rangle\langle \phi_{0k} |\msc{H}'|\nu_{0j}\rangle \delta(E(k)-\bar{E}) \,.
\end{align}
and $\bar{E}$ is the average energy of the neutrino beam. Within WW, Eq.~\eqref{Heff} appears naturally as a result of ``integrating out'' the new states, taking into account that the new states may be on-shell.
Moreover, it is easy to see that $\Gamma$ is positive definite. That is, WW only yields neutrino states that ``decay'' into the new states but never the other way around. Furthermore, off-diagonal $\Gamma_{ij}$ occur when different $\msc{H}_0$ eigenstates can access the same $\phi_{0k}$ state, i.e., the different neutrino ``mass'' eigenstates can ``decay'' into the same final state. 

In this work we assume that neutrino evolution is dictated by Eq.~\eqref{eff-eq}, but we are also be interested in violations of WW that may invalidate the constraint that $\Gamma$ is positive definite. In particular, these could happen if the matrix elements $\langle\phi_{0j}|\msc{H}'|\phi_{0k}\rangle$ cannot be neglected. In other words, we assume that there are conditions under which Eq.~\eqref{eff-eq} is a good description of neutrino propagation physics while the restriction that $\Gamma$ is positive definite need not apply. On the other hand, Eq.~(\ref{eq:Ple1}) is a  consequence of the more general hypothesis that the neutrinos mix with new, unidentified degrees of freedom, so we pay special attention to what these constraints imply. Of course, if one wishes to simply explore how well neutrino oscillations are governed by the standard laws of quantum mechanics, no constraints on $\Gamma$, other than those imposed by experimental data, need apply. 

Eq.~\eqref{eff-eq} can be written in the flavor basis, as in Eq.~(\ref{Scheq}), by performing the unitary transformation that links the two orthonormal sets of states $| \nu_{0i} \rangle$ and $| \nu_\alpha \rangle$. Solving Eq.~(\ref{Scheq}) is straightforward. Let $N$ be a \emph{generic} matrix such that
\be
\tilde{\msc{H}} = N\msc{H}N^{-1} \,,\quad \tilde{\msc{H}} = \trm{diag}\{h_1,\,h_2, \ldots \}, \label{NdefH}
\ee
where $h_i$ are complex numbers, and define the eigenstates $|\nu_i \rangle$ of the effective Hamiltonian as
\be
|\nu_{\alpha} \rangle = N_{\alpha i}|\nu_i \rangle. \label{Ndef}
\ee
The matrix $N$ is not uniquely defined by Eq.~\eqref{Ndef}; rescalings of the eigenvalues are still possible. In order to define it completely, we further impose 
\be
\langle \nu_i |\nu_i \rangle = 1, \forall i. \label{epn}
\ee
In general, the states $|\nu_i\rangle$ and  $|\nu_j\rangle$, $i\neq j$, are \emph{not} orthogonal. We define
\be
\langle \nu_i |\nu_j \rangle \equiv H_{ij} = (\mathbb{I}+\delta)_{ij} \label{delta-def}
\ee
where $\mathbb{I}$ is the identify matrix, $H$ and $\delta$ are Hermitian matrices, and $\delta_{ii}=0$, $\forall i$. It is convenient to express $N$ in terms of $\delta$. 
 Eq.~\eqref{flavor-ort} and \eqref{Ndef} imply
\be
N (\mathbb{I}+\delta^T)  N^\dagger = \mbb{I},
\ee
hence
\begin{eqnarray}
N\left(\mathbb{I}+\delta^T\right)^{1/2} &=& V, \\
N &=& V \left(\mathbb{I}+\delta^T \right)^{-1/2},
\end{eqnarray}
where $V$ is a unitary matrix. When $\delta=0$, $N$ is a unitary matrix and the Hamiltonian eigenstates form an orthonormal basis in spite of the fact that the $h_i$ are, in general, complex. This special case is the one usually considered when one addresses neutrino decay (see, for example, \cite{Frieman:1987as,Barger:1998xk}; for a detailed discussion see \cite{Lindner:2001fx}). It is equivalent to postulating that $M$ and $\Gamma$ can be simultaneously diagonalized, and the eigenvalues of $\Gamma$ are proportional to the lifetimes of the neutrino mass eigenstates. 

$\Gamma\ll M$ implies $\delta_{ij} \ll 1$ and 
\be
N \sim V \left(\mathbb{I}-\delta^T/2\right). \label{Napp}
\ee
We will restrict our discussions to this case, unless otherwise noted.

The solution to Eq.~(\ref{Scheq}), assuming that the neutrino is in state $|\nu_{\alpha}\rangle$ at $L=0$, is
\be
|\nu_\alpha(L) \rangle = \sum_iN_{\alpha i}e^{-i h_i L}(N^{-1})_{i\beta} |\nu_\beta\rangle \,,\label{nu(t)}
\ee
and the oscillation amplitudes are
\be
\mcl{A}_{\alpha\beta} = \langle \nu_\alpha |\nu_\beta(t) \rangle = N e^{-i \tilde{\msc{H}} L}N^{-1}, \label{amps}
\ee
trivially related to the oscillation ``probabilities,'' $P_{\alpha\beta}\equiv |\mcl{A}_{\alpha\beta}|^2$, which are the observables directly accessible to neutrino oscillation experiments. 

Eq.~(\ref{amps}) leads to unitarity-violating effects that  are qualitatively different from postulating the existence of new oscillation lengths (i.e., light sterile neutrinos), or postulating that the weak-interaction eigenstates are not orthogonal \cite{Lee:1977tib,Antusch:2006vwa}. Some of the differences are easy to spot. For instance Eq.~(\ref{amps}) does not allow for any flavor change in the limit $L\to 0$ ($NN^{-1}\equiv\mathbb{I}$ even if $N$ is not unitary!), unlike the effects discussed in \cite{Antusch:2006vwa}. Also, Eq.~(\ref{amps}) does not contain any new oscillation lengths: the new dimensionful parameters lead to exponential decay (or growth) of  $P_{\alpha\beta}$, as will be discussed in more detail in the next paragraphs. 

It is instructive to discuss the case of two neutrino flavors in detail in order to illustrate the phenomena described by Eq.~(\ref{amps}). In this case, we define
\be
\langle\nu_1|\nu_2 \rangle\equiv \epsilon e^{i\zeta},
\ee
where $\epsilon,\,\zeta$ are real and positive, $\zeta \in[0,2\pi)$. Further defining $h_{1,2}=a_{1,2}-ib_{1,2}$ and parameterizing the $2\times 2$ unitary matrix $V$ with the mixing angle $\theta$, in the usual way,\footnote{It is straightforward to show that, like in the standard unitary-case, potential ``Majorana phases'' in $V$ play no role in neutrino oscillations, even for $\delta\neq 0$.} we find
\begin{equation}
N\sim \left(\begin{array}{cc} \cos\theta & \sin\theta \\ -\sin\theta & \cos\theta \end{array}\right)
\left(\begin{array}{cc} 1 & -\frac{1}{2}\epsilon e^{-i\zeta} \\ -\frac{1}{2}\epsilon e^{i\zeta} & 1 \end{array}\right),
\end{equation}
keeping in mind that $\epsilon\ll 1$. Setting $\alpha,\beta = e,\mu$ for the sake of definiteness we find the oscillation probabilities
\begin{eqnarray}
 P_{ee} &=& e^{-2b_1 L}\cos^4\theta + e^{-2b_2 L}\sin^4\theta + \frac{1}{2}e^{-(b_1+b_2)L} 
\sin2\theta\left[ \sin2\theta\cos \Delta L - 2\epsilon \sin\zeta \sin\Delta L \right], \\
 P_{e\mu} &=& \frac{1}{4}\left( e^{-2b_2L} + e^{-2b_1 L} - 2e^{-(b_1+b_2)L}\cos\Delta L\right)
\sin2\theta(\sin2\theta - 2\epsilon\cos\zeta), \\
 P_{\mu e} &=& \frac{1}{4}\left( e^{-2b_2L} + e^{-2b_1 L} - 2e^{-(b_1+b_2)L}\cos\Delta L\right)
\sin2\theta(\sin2\theta + 2\epsilon\cos\zeta), \\
 P_{\mu\mu} &=&  e^{-2b_2 L}\cos^4\theta + e^{-2b_1 L}\sin^4\theta + \frac{1}{2}e^{-(b_1+b_2)L}
\sin2\theta\left[ \sin2\theta\cos \Delta L + 2\epsilon \sin\zeta \sin\Delta L \right],
\end{eqnarray}
where $\Delta = a_2- a_1$ plays the role of $\Delta m^2/2E$ in the standard case and can be chosen positive. The expressions above ignore terms of ${\cal O}(\epsilon^2)$, an approximation that is not appropriate in the limit $\theta\to 0$. The two new dimensionful parameters $b_{1,2}$ lead to the exponential decay/growth of all oscillations probabilities. Lorentz invariance dictates that $b_{i}\propto d_{i}/E$, where $d_{1,2}$ are constants with dimensions of mass-squared. In the limit $\epsilon\to0$ we recover the well-known expressions for neutrino oscillations under the assumption that the neutrino mass eigenstates have a finite lifetime. In the more general case where the ``mass'' eigenstates do not coincide with the ``decay'' eigenstates, $\epsilon\neq 0$ and the non-unitarity of the propagation leads to  new ``mixing'' parameters, $\epsilon$ and $\zeta$. 

As discussed earlier, the physics responsible for $\Gamma\neq 0$ imposes constraints on the different parameters. In the two-flavor case, in the basis where $M$ is diagonal with diagonal elements $a_1$ and $a_2$ (chosen positive), $\Gamma_{ii}=b_i$ in the limit $\Gamma_{ij}\ll a_1,a_2$, $\forall i,j$. In the same basis, defining $\Gamma_{12}=\Gamma_{21}^*=b$, 
\begin{equation}
\epsilon e^{i\zeta} = -\frac{2ib}{\Delta}.
\end{equation}
In the context of the Weisskopf-Wigner approximation, $\Gamma$ is constrained to be positive-definite: $b_1,b_2>0$, $b_1b_2\ge |b|^2$. In turn, these imply that $\epsilon\le 2\sqrt{b_1b_2}/\Delta$.

The less stringent constraint $\sum_{\beta}P_{\alpha\beta}\le 1$, for all $\alpha$, translates into $b_1, b_2\ge0$ and 
\begin{equation}
\epsilon\le C_{\zeta}\frac{b_1\cot\theta+b_2\tan\theta}{\Delta} \cap \epsilon\le C_{\zeta}'\frac{b_2\cot\theta+b_1\tan\theta}{\Delta},
\end{equation}
where $C_{\zeta}, C_{\zeta}'$ are ${\cal O}(1)$  non-illuminating functions of $\zeta$ such that $1< C_{\zeta} \lesssim 2$. This constraint allows for $\epsilon\neq 0$ as long as both $b_1,b_2\neq 0$. Importantly and opposed to the case in which WW is assumed, we find that for small mixing ($\sin2\theta\ll 1$), $\epsilon$ values larger than $b_{1,2}/\Delta$ are allowed. This is potentially relevant for the ``$1-3$ sector,'' as well as for the application of this formalism to sterile neutrinos.

Regardless of the origin of $\Gamma$, it is instructive to consider the case $|b_1 L|,|b_2 L|\to 0$. Under these circumstances 
\begin{eqnarray}
P_{ee} &=& 1- \sin^22\theta\sin^2(\Delta L/2)
-\sin2\theta\left(\epsilon \sin\zeta\right) \sin\Delta L, \\
P_{e\mu} &=& \sin2\theta(\sin2\theta - 2\epsilon\cos\zeta)\sin^2(\Delta L/2),  \\
P_{\mu e} &=& \sin2\theta(\sin2\theta + 2\epsilon\cos\zeta)\sin^2(\Delta L/2), \\
P_{\mu\mu} &=&  1- \sin^22\theta\sin^2(\Delta L/2)  
+\sin2\theta\left(\epsilon \sin\zeta\right) \sin\Delta L.
\end{eqnarray}
Note that these are only good approximations in scenarios where $\sum_{\beta}P_{\alpha\beta}> 1$, in which case $P_{\alpha\beta}$ need to be carefully reinterpreted as they cannot stand, mathematically speaking, for probabilities. Nonetheless, the above expressions are easy to explore -- only two new dimensionless parameters appear -- and are useful in order to illustrate the consequences of $\epsilon,\zeta\neq0$, 

Some interesting features are worthy of note. Even when the decay effects are ``turned off,'' unitarity is violated -- $P_{ee}+P_{e\mu}\neq 1$ -- along with, in the case $\zeta\neq \pi/2,3\pi/2$, time-reversal invariance -- $P_{e\mu}\neq P_{\mu e}$. The oscillation length is the same for all $P_{\alpha\beta}$, $L_{\rm osc}=2\pi/\Delta$, but the survival probabilities are ``out of phase'', i.e., maxima and minima do not correspond to $L=n L_{\rm osc}$, for natural $n$. The amplitudes of the oscillations -- differences between the smallest and largest $P_{\alpha\beta}$ -- for appearance and disappearance are also different. For example
\begin{eqnarray}
A_{ee} &=& \sin^22\theta\sqrt{1+\left(\frac{2\epsilon\sin\zeta}{\sin2\theta}\right)^2},\\
A_{\mu e} &=& \sin^22\theta\left(1+\frac{2\epsilon\cos\zeta}{\sin2\theta}\right).
\end{eqnarray}
A measurement of $\nu_e$ disappearance can report an effective mixing angle $\sin^22\theta_{\rm eff}\equiv A_{\alpha\beta}$ that is different from that observed in appearance experiments.  For example, if $2\epsilon\ll \sin2\theta$, non-unitarity effects in disappearance are much smaller than those in appearance, unless $\cos\zeta$ is very small. 

Under a CP-transformation $N\to (N^{-1})^T$, so antineutrinos are governed by the same differential equation except for $\zeta\to\pi-\zeta$, i.e., $\sin\zeta\to\sin\zeta$, $\cos\zeta\to-\cos\zeta$. For example, ignoring term of ${\cal O}(\epsilon^2)$,  
\begin{eqnarray}
& P_{\bar{e}\bar{e}} = e^{-2b_1 L}\cos^4\theta + e^{-2b_2 L}\sin^4\theta + \frac{1}{2}e^{-(b_1+b_2)L}
\sin2\theta\left[ \sin2\theta\cos \Delta L - 2\epsilon \sin\zeta \sin\Delta L \right] \,, \\
& P_{\bar{\mu} \bar{e}} = \frac{1}{4}\left( e^{-2b_2L} + e^{-2b_1 L} - 2e^{-(b_1+b_2)L}\cos\Delta L\right)
\sin2\theta(\sin2\theta - 2\epsilon\cos\zeta) \,. 
\end{eqnarray}
As expected, CPT-invariance is preserved, i.e., $P_{\alpha\beta}=P_{\bar{\beta}\bar{\alpha}}$, while CP-invariance is not unless $\zeta=\pi/2,3\pi/2$. Non-unitary propagation leads to new CP-invariance-violating phenomena, even in the two-flavor case.

CP-invariance and T-invariance violation are also present, as long as $\epsilon\cos\zeta\neq0$, in the regime where the oscillatory terms average out, i.e., $\Delta L\gg 1$. In this case,
\begin{eqnarray}
P_{ee}&=& e^{-2b_1 L}\cos^4\theta + e^{-2b_2 L}\sin^4\theta, \\
P_{\mu\mu}&=& e^{-2b_1 L}\sin^4\theta + e^{-2b_2 L}\cos^4\theta, \\
P_{e\mu} &=& \frac{\sin^22\theta}{4}\left( e^{-2b_2L} + e^{-2b_1 L}\right)\left(1 - \frac{2\epsilon\cos\zeta}{\sin2\theta}\right),  \\
P_{\mu e} &=& \frac{\sin^22\theta}{4}\left( e^{-2b_2L} + e^{-2b_1 L}\right)\left(1 + \frac{2\epsilon\cos\zeta}{\sin2\theta}\right). \, 
\end{eqnarray}
$P_{\bar{e}\bar{e}}=P_{ee}$, while $P_{e\mu}-P_{\mu e}=P_{e\mu}-P_{\bar{e}\bar{\mu}}\propto\epsilon\cos\zeta$.

All expressions above ignore terms ${\cal O}(\epsilon^2)$ and are not good approximations in the limit $\sin2\theta\ll \epsilon$. In the limit $\theta\to 0$, it is easy to compute the oscillation probabilities. Ignoring ${\cal O}(\epsilon^3)$  terms\footnote{Here we include ${\cal O}(\delta^2)$ contributions to $N$, replacing Eq.~(\ref{Napp}) with $N \sim\mathbb{I}(1+3\epsilon^2/8)-\delta/2.$}
\begin{eqnarray}
P_{ee}&=&  \left(1+\frac{\epsilon^2}{2}\right)e^{-2b_1L}-\frac{\epsilon^2}{2}e^{-(b_1+b_2)L}\cos\Delta L, \\
P_{\mu\mu}&=& \left(1+\frac{\epsilon^2}{2}\right)e^{-2b_2L}-\frac{\epsilon^2}{2}e^{-(b_1+b_2)L}\cos\Delta L,  \\
P_{e\mu} &=& \frac{\epsilon^2}{4}\left(e^{-2b_1L}+e^{-2b_2L}-2 e^{-(b_1+b_2)L}\cos\Delta L \right) \,, 
\end{eqnarray}
while $P_{\mu e}=P_{e\mu}$. CP-invariance is preserved -- none of the expressions depend of $\zeta$, so $P_{\alpha\beta}=P_{\bar{\alpha}\bar{\beta}}$, $\forall \alpha,\beta$, but nonzero $\epsilon$ implies flavor change as long as the neutrino masses are different ($\Delta\neq0$). 

In the $\theta\to 0$ limit, the constraint $\sum_{\alpha}P_{\alpha\beta}<1$ translates into $b_1,b_2\ge0$ and
\begin{equation}
\epsilon\le \sqrt{\pi\frac{{\rm min}(b_1,b_2)}{2\Delta}},
\end{equation}
where ${\rm min}(b_1,b_2)$ indicates the smaller between $b_1$ and $b_2$, so nonzero $\epsilon$ requires both $b_1,b_2$ nonzero. Furthermore, $\sum_{\alpha}P_{\alpha\beta}<1$ also implies that oscillatory effects cannot dominate over the exponential decay.

It is easy to extend the discussion to the three-flavor case. We define
\be
\langle \nu_1 |\nu_2 \rangle = \epsilon_{3} e^{i\zeta_{3}} \,,\quad
\langle \nu_1 |\nu_3 \rangle = \epsilon_{2} e^{i\zeta_{2}} \,,\quad
\langle \nu_2 |\nu_3 \rangle = \epsilon_{1} e^{i\zeta_{1}} \,
\ee
where $\epsilon_i\ge 0$ and $\zeta_i\in[0,2\pi)$. Thus, in the limit $\epsilon_i\ll 1$, $N$ is given by Eq.~(\ref{Napp}) where 
\begin{equation}
\delta=\left(
\begin{array}{c c c}
 0 & e^{-i \zeta_{3}} \epsilon_{3} & e^{-i \zeta_{2}} \epsilon_{2} \\
 e^{i \zeta_{3}} \epsilon_{3} & 0 & e^{-i \zeta_{1}} \epsilon_{1} \\
 e^{i\zeta_2} \epsilon_{2} & e^{i \zeta_{1}} \epsilon_{1} & 0
\end{array}
\right) ,
\label{k3}
\end{equation}
while the $3\times 3$ unitary matrix $V$ can be parameterized in the usual way, 
\be
V=\left(\begin{array}{c c c}
c_{12} c_{13} & s_{12} c_{13} & s_{13} e^{-i \delta_{\rm CP}} \\
-s_{12} c_{23} - c_{12} s_{23} s_{13} e^{i \delta_{\rm CP}} & c_{12} c_{23} - s_{12} s_{23} s_{13} e^{i \delta_{\rm CP}} & s_{23} c_{13} \\
s_{12} s_{23} - c_{12} c_{23} s_{13} e^{i \delta_{\rm CP}} & -c_{12} s_{23} - s_{12} c_{23} s_{13} e^{i \delta_{\rm CP}} &  c_{23} c_{13}  \end{array} \right), 
\label{um}
\end{equation}
where $s_{ij}=\sin{\theta_{ij}}$, $c_{ij}=\cos{\theta_{ij}}$ and $\delta_{\rm CP}$ is the ``Dirac'' CP-odd phase. As in the two-flavor case, neutrino oscillations are not sensitive to ``Majorana phases'' in $V$.

Using Eq.~(\ref{amps}), the transition probabilities can be written as
\begin{eqnarray}
P_{\alpha \beta}&=& |e^{-i h_{i} t}|^{2} (N_{\alpha i}N^{-1}_{i \beta })(N^{\ast}_{\alpha i}{N}^{-1\ast}_{i \beta })\nonumber\\ & &+\sum_{i\neq k} e^{-i h_{i} t}e^{i h_{k} t}(N_{\alpha i}N^{-1}_{i \beta })(N^{\ast}_{\alpha k}N^{-1\ast}_{k \beta }),
\label{p3fg}
\end{eqnarray}  
where $h_i=a_i-ib_i$. As before, $\Delta_{ij}=(a_i-a_j)$ play the role of oscillation frequencies, $\Delta m^2_{ij}/2E$, while the different $b_{i}\propto d_i/E$, $i=1,2,3$ lead to exponential decay (or growth).  

Complete expressions for the different $P_{\alpha\beta}$ are rather cumbersome and not particularly illuminating. They depend on all decay parameters as well as the new mixing parameters $\epsilon_{1,2,3}$ and the new CP-odd phases $\zeta_{1,2,3}$. As in the two-flavor case, the transitions for antineutrinos are governed by the same expressions, except for $N\to(N^{-1})^T$, or $V\to V^*$, $\Delta\to -\Delta$. In terms of the parameterization introduced here, this translates into $\delta_{\rm CP}\to-\delta_{\rm CP}$, $\zeta_i\to\pi-\zeta_i$. As expected, CPT-invariance is preserved, $P_{\alpha\beta} = P_{\bar{\beta}\bar{\alpha}}$, but CP-invariance and T-invariance are violated unless $\epsilon_i\cos\zeta_i=0$ for all $i$. This is true even when three-flavor effects are ``turned-off'' (some mixing angles and some mass-squared differences vanish) and persists in the limit  $\Delta_{31}L,\Delta_{21}L\gg 1$. 

For the case of standard neutrinos, bounds on some of the new physics parameters have been discussed in the literature, mostly in the context of ``standard'' neutrino decay ($\delta_{ij}=0$ in Eq.\eqref{delta-def}). However, the bounds on the $\epsilon_i$ we will discuss are indirect, all of them deriving from the bounds on the $b_i$, and relatively weak if WW is abandonned. Here we summarize qualitatively some of these bounds. A quantitative discussion on how non-unitary neutrino propagation is constrained by current data and the potential reach of next-generation experiments is currently being pursued in the follow-up to this letter \cite{new}. 

In what follows, it will prove convenient to define the Hamiltonian eigenstates as 
\begin{equation}
h_i=a_i-ib_i\equiv \frac{1}{2E}\left(m_i^2 - id_i\right),
\end{equation}
where the $d_i$'s have dimensions of mass-squared. Below, we assume that $m_i^2$ agree with the results obtained by analyzing the world's neutrino data under the assumption that neutrino propagation is unitary, and use the standard definition for the ordering of the masses \cite{deGouvea:2013onf}.

A finite neutrino lifetime can dramatically impact all the indirect information we have on primordial neutrinos. The nature of the decay products and the interactions, however, plays a role in determining if and how decaying neutrinos impact the cosmic microwave background, structure formation, etc, so bounds on decay parameters are very model-dependent (see, for example, \cite{Beacom:2004yd,Hannestad:2005ex}), ranging from very stringent to non-existent. A future observation of nonzero neutrino mass effects in cosmic surveys may change the picture dramatically \cite{Serpico:2007pt}. 

The recent observation of ultra-high energy neutrinos from potentially extra-galactic sources \cite{Aartsen:2013bka} implies that at least one of the $d_i$'s is tiny.  Much more information can be obtained -- especially if the $d_i$'s are not zero -- with more statistics and flavor information \cite{Beacom:2002vi} (see also, for example, \cite{Meloni:2006gv,Baerwald:2012kc,Pakvasa:2012db,Dorame:2013lka,Fu:2014gja}). A similarly very stringent bound -- at least one of the $d_i$ is zero for all practical purposes -- comes from the observation of neutrinos from SN1987A \cite{Frieman:1987as}. 

Closer to home, solar neutrino data place strong constraints on some neutrino decay parameters. Data on $^8$B-neutrinos from Super-Kamiokande and SNO constrain $d_2\lesssim 10^{-11}$~eV$^2$ \cite{Beacom:2002cb}, assuming $d_1$ is zero. A more detailed analysis \cite{new}, including recent data from Borexino, should also allow one place bounds on $d_1$. Atmospheric neutrino data constrain $d_3\lesssim 10^{-5}$~eV$^2$ \cite{GonzalezGarcia:2008ru}, assuming $d_{1,2}$ are zero.\footnote{A new analysis of recent long-baseline neutrino data quotes a new bound, which is about an order of magnitude stronger \cite{verynew}. This analysis appeared on the preprint archives as this manuscript was about to be submitted.} Shorter-baseline neutrino experiments (e.g., reactor and beam experiments) constrain $d_i\lesssim 10^{(-7)-(-4)}$~eV$^2$ for all $i=1,2,3$.

If $\Gamma$ is positive definite, $\epsilon_i\lesssim \sqrt{d_jd_k}/\Delta m^2$ and the bounds above imply that at least two of the $\epsilon_i$ are tiny given the constraints from solar neutrino data (at least $\epsilon\lesssim 10^{-5}$), while the third one might be of order a few percent ($\lesssim 10^{-5}$~eV$^2/|\Delta m^2_{13}|$). The less stringent requirement $\sum_{\alpha}P_{\alpha\beta}\le 1$ allows for more $\epsilon_i$ of order several percent, potentially enhanced by the fact that $\theta_{13}$ is small. A detailed discussion will be presented in \cite{new}. If no restriction is imposed on $\Gamma$ or $\sum_{\alpha}P_{\alpha\beta}$, joint analyses of nonzero $d_i$ and $\epsilon_i$ are required in order to establish the currently allowed values of these phenomenological parameters. 

In summary, non-unitary neutrino propagation can be realized if neutrinos couple to new very light states -- lighter than the active neutrinos -- which can interact among themselves. Here, we study some of the consequences of this hypothesis. This scenario is qualitatively different from the unitarity violation setups that have been prevously discussed in the literature and leads to new phenomena -- including new sources of CP-invariance violation and new mass-scales -- that can only be probed in oscillation experiments. A lot of work remains to be done, including a quantitative discussion of current bounds on non-unitary propagation, three-flavor phenomenology and prospects for next-generation oscillation experiments, potential applications to the short-baseline anomalies and other searches for new neutrino oscillation lengths, etc. We also need to address matter effects, especially when it comes to addressing their impact on the non-unitarity parameters -- especially the $\epsilon$ parameters -- and some of the constraints we have discussed here.  

\section*{Acknowledgements}

The work of JMB, AdG, and DH is sponsored in part by the DOE grant \#DE-FG02-91ER40684. The work of RLNO is supported by grant 2013/11651-2 from FAPESP (Brazil). D.H. is indebted to E. Fernandez-Martinez, B. Gavela and A. Yu. Smirnov for illuminating discussions. He is also grateful for the hospitality of the Instituto de Física Teórica (Madrid) and of MPIK (Heidelberg) where part of this work was developed.

\end{document}